\documentclass[]{aastex631}

\usepackage{amsmath}
\usepackage{amssymb}
\usepackage{comment}
\usepackage[T1]{fontenc}
\usepackage{graphicx}
\usepackage{booktabs}
\usepackage{dcolumn}
\usepackage{bm}
\usepackage{hyperref}
\usepackage{ragged2e}
\usepackage{academicons}
\usepackage{xcolor}
\newcommand{\cmmnt}[1]{\ignorespaces}

\def\rg{r_{\rm g}}
\def\tg{t_{\rm g}}
\def\rB{r_{\rm B}}
\def\Ra{R_{\rm HL}}
\def\mel{m_{{\rm e}}}
\def\mp{m_{\rm p}}
\def\nel{n_{{\rm e}^-}}
\def\np{n_{\rm p}}
\def\Thetae{\Theta_{\rm e}}
\def\Gammaeq{\Gamma_{\rm eq}}
\def\rin{r_{\rm in}}
\def\rou{r_{\rm out}}
\shorttitle{BHL accretion onto isolated black holes}
\shortauthors{Tripathi et al.}

\begin{document}

\title{On Disk Formation around Isolated Black Holes via Stream Accretion}
\correspondingauthor{Indranil Chattopadhyay}
\email{indra@aries.res.in}

\author[0009-0002-7498-6899]{Priyesh Kumar Tripathi}
 \affiliation{Aryabhatta Research Institute of Observational Sciences (ARIES),\\ Manora Peak, Nainital, 263001, India\\}
 \affiliation{Department of Applied Physics, Mahatma Jyotiba Phule Rohilkhand University,\\ Bareilly, 243006, India\\}

\author[0000-0002-2133-9324]{Indranil Chattopadhyay}
 \affiliation{Aryabhatta Research Institute of Observational Sciences (ARIES),\\ Manora Peak, Nainital, 263001, India\\}

\author[0000-0002-9036-681X]{Raj Kishor Joshi}
 \affiliation{Aryabhatta Research Institute of Observational Sciences (ARIES),\\ Manora Peak, Nainital, 263001, India\\}



\begin{abstract}
  We investigate accretion onto an isolated black hole from uniform winds. If the winds are directed towards the black hole, then the accretion process can be well described by the classical Bondi-Hoyle Lyttleton or BHL accretion. If the wind is not directed towards the black hole and flows past it, then a smaller fraction of the flow can be attracted by the black hole, and this type of accretion cannot be described by the classical BHL, and we coin the second kind as the lateral BHL. We show that the classical BHL cannot form an accretion disk, while lateral BHL can form transient accretion disks. To describe the thermodynamics of the flow, we have used a variable adiabatic index equation of state which depends on the temperature of the flow as well as the composition of the gas. We show that the electron-proton gas forms an accretion disk, which disappears and forms a shock cone, only to form the disk again at a later time, while for flows with less protons, the accretion disk, once lost, does not reappear again. Only when the flow is pair-dominated does it form a persistent accretion disk. We also show that a shock cone is less luminous than the accretion disk.
\end{abstract}





\section{\label{sec:Introduction}Introduction}
   Accretion onto isolated black holes is gaining importance after many such black holes have been detected by the LIGO experiments. The difference between the accretion onto an isolated black hole and a black hole with a companion is the mass/energy supply. In the case of a black hole with a companion star, the mass supply will be dictated by the state of the companion. An isolated black hole can accrete matter if it plows through a diffused gas or if it is impacted by wind or a flux of matter onto it.
   The accretion process in which an isolated black hole accretes matter by plying through a gaseous medium is known as the Bondi–Hoyle–Lyttleton (BHL) accretion \citep{1939_hoyle_lyttleton_PCPS...35..405H,1944_bondi_hoyle_MNRAS.104..273B,1952_bondi_MNRAS.112..195B}.
   If a compact object plies through a diffused gas, then from the reference frame of the compact object, the gas material of the cloud is streaming past it at the same relative speed. So an isolated black hole traveling through a gas is equivalent to a uniform wind impinging on the black hole. Therefore, both can be studied under the guise of the BHL process.
   However, if the wind is not directed towards the black hole, then the bulk of the matter will be streaming past the black hole, and only a small fraction will be attracted by gravity towards the black hole. The second case cannot be studied as the classical BHL model, and we coin the second case as the lateral BHL model.   
   Incidentally, the BHL model of accretion has been applied to a variety of problems \citep{2004_Edgar_NewAR..48..843E}. For example,
   it was used as a first approximation of an accretion model of a high-velocity wind in binary systems \citep{1978ApJ_Petterson...224..625P}; or, study of accretion flow in young stellar clusters \citep{2001_Bonnell_et_al_MNRAS.323..785B}; or, central black hole (BH) accreting material from an expanding supernova shock etc. 
   \citet{2002_Pfahl_et_al_ApJ...571L..37P} studied wind accretion onto a neutron star, although the presence of magnetic field may reduce the accretion rate \citep{2001_Toropina_ApJ...561..964T}.
   There are a number of studies where BHL accretion has been invoked to study the accretion of ambient matter onto an isolated black hole traveling through it \citep{2002_Agol_Kamionkowski_MNRAS,2013_Fender_etal_MNRAS,2019_Tsuna_Kawanaka_MNRAS,2021_Scarcell_etal_MNRAS.505.4036S}.
 
   Purely 3D hydrodynamical simulation of BHL accretion in the Newtonian case was studied in a series of papers by \citet{1994a_Ruffert_A&AS..106..505R, 1994b_Ruffert_ApJ...427..342R, 1994_Ruffert_Arnett_ApJ...427..351R} using ideal gas approximation. \citet{2015_Lora-Clavijo_ApJS..219...30L} studied the relativistic BHL accretion with density gradients in the ambient medium. \citet{2017_Cruz_Osorio_MNRAS.471.3127C} studied the relativistic BHL accretion in the presence of randomly distributed small rigid bodies around BH, which represent the substructures like stars passing close to BH, which causes some variability in accretion rates. \citet{2014_Lee_ApJ...783...50L} studied the BHL accretion in an isothermal magnetized plasma and showed that the magnetic fields reduce the accretion rates.
   Recently, 3D GRMHD simulations of Bondi–Hoyle–Lyttleton Accretion \citep{2023_Kaaz_ApJ...950...31K} showed that the jet could be launched by accretion onto rapidly rotating BHs moving through a magnetized gaseous medium.
   \citet{2020_Li_MNRAS.494.2327L, 2020_Valenti_A&A...638A.132B, 2022_Valenti_A&A...660A...5B} estimated the impact of mechanical feedback on Bondi-Hoyle-Lyttleton accretion showing that accretion rate is significantly reduced by the production of outflows in the medium.
   {Disk formation in BHL accretion has been investigated by \citet{2013_Blondin_ApJ...767..135B,2019_Mellah_et_al_A&A...622A.189E,2019_xu_stone_MNRAS.488.5162X}. \citet{2015_MacLeod_Ramirez-Ruiz_ApJ...803...41M,2017_MacLeod_et_al_ApJ...838...56M,2017_Murguia-Berthier_et_al_ApJ...845..173M} studied accretion disk assembly in BHL-like flows around objects embedded within a common envelope.}
   
   The aim of this work is to investigate and compare the accretion processes of (a) supersonic wind directed towards a BH with (b) a wind where the BH is not in its line of propagation. The outflow or wind can be imagined to originate from an accidental neighborhood star. 
   Far from the BH, the flow should be cold (adiabatic index $\Gamma \sim 5/3$), but closer to the BH, it would be much hotter. To capture this physics, we have employed a variable $\Gamma$ equation of state (EoS). Moreover, the `hotness' of a medium does not depend only on the absolute value of temperature but also on the ratio of the inertia and the random kinetic energy of the constituent particle of the gas. As a result we also investigate whether composition has any effect on the accretion or not.

   The outline of the paper is as follows. In section~\ref{sec:Theoretical background}, we describe the background of the theoretical development of BHL accretion. Section~\ref{sec:Assumptions and governing equations} describes the governing equations of our numerical approach and the assumptions used, along with the need for relativistically correct EoS. Then, we describe the setup of our numerical model and the parameters used in section~\ref{sec:Numerical Model and Setup}. Finally, we present our results in section~\ref{sec:Results}, followed by the summary and discussions in section~\ref{sec:Summary and Discussions}.

\section{\label{sec:Theoretical background}Theoretical background}
   \begin{figure*}
    \begin{center}
      \includegraphics[width=\textwidth]{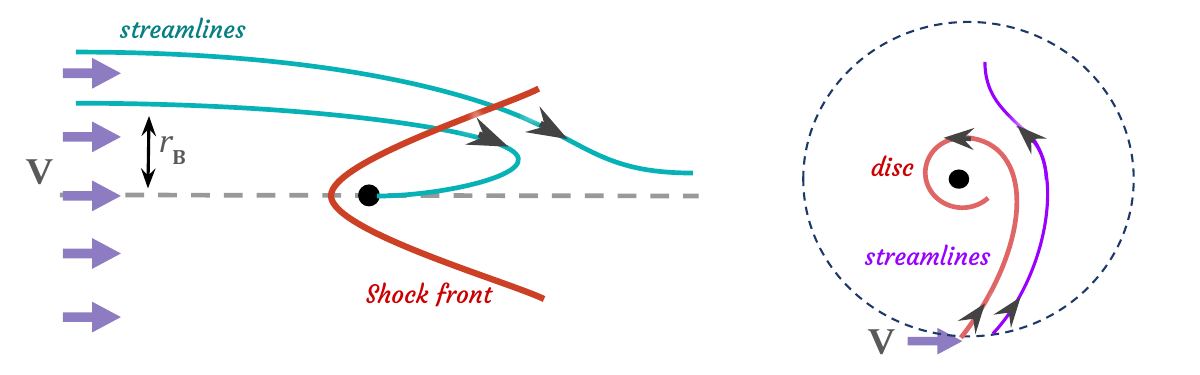}
    \end{center}
       \caption{Left: Geometry of the classical BHL accretion onto an isolated BH. The thick red line in the left panel is the shock front.
       Right: Geometry of the lateral BHL. The capture streamline (red) and the streamline which leaves the domain (violet) are shown.}
       \label{fig:BHL accretion geometry}
   \end{figure*}
  The first model (left panel of Figure~\ref{fig:BHL accretion geometry}) is the classical BHL model. A supersonic flow directed towards the gravitational center would be accreted if the streamline is within a length scale in which the gravitational pull is dominant over all other energies. 
  Such a length scale is given by,
      \begin{equation}
          \Ra = \frac{2GM}{v_\infty^2}
      \end{equation}
   where $v_\infty$ is the relative velocity of the flow with respect to the accretor of mass $M$. The accretion rate of the system is determined by the mass flux of the material having an impact parameter smaller than the $\Ra$
   \citep{1939_hoyle_lyttleton_PCPS...35..405H}, 
      \begin{equation}
          \label{eq: HL accretion rate}
          \begin{split}
              \dot{M}_{\rm HL} &= {\pi {\Ra}^2} {\rho_\infty} {v_\infty} = 4 \pi (GM)^2 \frac{\rho_\infty}{v_\infty^3} \\ 
           & \approx {7.21 \times 10^{-29}} \frac{\tilde{M}^2 \tilde{n}}{\tilde{v_\infty}^3} M_\odot / year
          \end{split}
      \end{equation}
   where $\rho_\infty$ is the density of the undisturbed flow at infinity and $\dot{M}_{\rm HL}$ is the Hoyle-Lyttleton accretion rate. $\tilde{M}$, $\tilde{n}$, $\tilde{v_\infty}$ are the normalized mass of the accretor in units of the solar mass ($M_\odot$), the normalized number density in units of atoms per cubic centimeter, and the normalized relative velocity in units of speed of light. For an accretor of mass $1 M_\odot$ moving with $0.01c$ velocity in the medium having density 1 H-atom/cc, the Hoyle-Lyttleton accretion rate would be around $10^{-23} M_\odot/$year. A more general formula involving the pressure effects of the fluid was given
   by \citet{1952_bondi_MNRAS.112..195B},
      \begin{equation}
          \label{eq: BH accretion rate}
          \dot{M}_{\rm BH} = 4 \pi (GM)^2 \frac{\rho_\infty}{(v_\infty^2 + {c}_{\rm s, \infty}^2)^{3/2}}
      \end{equation}
   where $c_{\rm s,\infty}$ is the sound speed of the flow at infinity and $\dot{M}_{\rm BH}$ is known as the Bondi-Hoyle accretion rate. These two accretion rates are related as,
      \begin{equation}
          \dot{M}_{\rm BH} = \dot{M}_{\rm HL} \left( \frac{{\mathcal{M}_\infty}^2}{{\mathcal{M}_\infty}^2 + 1} \right) ^{3/2}
      \end{equation}
   where $\mathcal{M}_\infty = v_\infty/c_{s,\infty}$ is the Mach number of the incoming flow. For very high Mach numbers, $\mathcal{M}_\infty >> 1$, when the pressure effects are negligible compared to the kinematical effects, the two accretion rates are the same.
   Following the above, one can define a Bondi radius, which is a length scale where the gravitational potential is equal to the injected kinetic and thermal energy defined as
   \begin{equation}
   \label{eq:bondi_rad}
   \rB = \frac{GM}{0.5 v^2_\infty+h_{\infty}},
   \end{equation}
   In the above, $h_\infty$ is the specific enthalpy of the flow at the injection.
   This gives a very crude estimate of the sphere of influence of the gravity of the black hole.

   In the case of lateral BHL, i.e., the flow is initially not directed towards the black hole, but a fraction of matter, with an impact parameter less than the $\rB$ are expected to be gravitationally captured.
   In the right panel of Figure~\ref{fig:BHL accretion geometry}, the cartoon diagram of lateral BHL is presented.

\section{\label{sec:Assumptions and governing equations}Assumptions and governing equations}
   We perform numerical simulations of the Bondi-Hoyle-Lyttleton accretion flow, which is described by the time-dependent hydrodynamical equations of motion. The coordinate-free conservative form of these equations for non-relativistic inviscid flow is:
      \begin{subequations}
      \label{eq:HD eqns}
        \begin{align}
          & \frac{\partial \rho}{\partial t} + 
                            \mathbf{\nabla} \cdotp (\rho \mathbf{v}) = 0 
      \\  & \frac{\partial (\rho \mathbf{v})}{\partial t} + 
                       \mathbf{\nabla} \cdotp (\rho\mathbf{v \otimes v} + p_t\mathbf{I}) = \rho \mathbf{g}
      \\  & \frac{\partial E}{\partial t} + 
                 \mathbf{\nabla} \cdotp \left[(E+p_t)\mathbf{v}\right] = \rho \mathbf{v \cdotp g} ,
        \end{align}
      \end{subequations} 
  where $\rho, \mathbf{v} $ and $p_t$ denote the mass density, bulk velocity, and thermal pressure, respectively, $\mathbf{I}$ is a unit tensor, and $E=e+\rho v^2 /2$ is the total energy density or the sum of thermal energy density and the kinetic energy density of the fluid element.
  The gravitational acceleration 
  $\mathbf{g} = -\mathbf{\nabla} \phi$, with $\phi$ being the gravitational potential of the source. To take care of strong gravity near the compact object, we have assumed the \citet{1980PWPA&A....88...23P} potential, which closely mimics
  the effects of strong gravity near a Schwarzschild BH. For an accreting object of mass M having no self-gravity, the form of this potential is:
     \begin{equation}
         \phi (r) = -\frac{GM}{R-\rg}
     \end{equation}
  where, $R=\sqrt{r^2+z^2}$ and $\rg$ is the Schwarzschild radius of the central object.
  
\subsection{\label{sec:EOS}Equation of state}
   To close the system of equations \eqref{eq:HD eqns}, we additionally need an equation of state (EoS), which carries all the microphysical information through the thermodynamic properties of the fluid. To account for the transition between very high to very low temperatures in flow, one needs to use the EoS, which carries temperature information. The first relativistically correct EoS given by \citet{Chandrasekhar1939} is difficult to use numerically due to the involvement of the modified Bessel functions. Further, the composition is also important for a flow to be thermally relativistic or non-relativistic depending on whether the thermal energy $(kT)$ is comparable to its rest mass energy $(mc^2)$. An approximate but relativistically correct equation of state was given by \citet{2009Chattopadhyay&RyuApJ} (CR EoS) with variable $\Gamma$ for multispecies flows, where the internal energy in the unit system $c=1$ is given as \citep{2021joshietal_MNRAS.502.5227J};
      \begin{equation}
             e = \rho f 
      \end{equation}
    where, 
       \begin{equation}
          f = 1 + (2-\xi)\Theta \left[\frac{9\Theta+6/\tau}{6\Theta+8/\tau}\right] + 
            \xi \Theta \left[\frac{9\Theta+6/\eta \tau}{6\Theta+8/\eta \tau}\right]
        \label{eq:CR EoS}
       \end{equation}
     In the above equation, $\xi = \np/\nel$ and $\eta = \mel/\mp$ are ratios of the 
     number density of the proton to electron and mass of the electron to the proton, respectively. 
     In addition, $\tau = 2-\xi + \xi /\eta$ , which is a function of the composition ($\xi$) of the flow. 
     $\Theta = p/\rho$ is a measure of temperature (in physical units, it is the ratio of thermal energy with the rest mass energy or $kT/ (\mel \tau c^2)$). 
     The specific enthalpy is given as ; 
        \begin{equation}
            h = (e + p)/\rho = f + \Theta
        \end{equation}   
     The polytropic index N is given as ; 
        \begin{equation} 
         \begin{split}
             N   = \rho \frac{\partial{h}}{\partial{p}} - 1 
               & = \frac{\partial{f}}{\partial{\Theta}} 
                 = 6 (2-\xi) \left[ \frac{9\Theta^2+24\Theta/\tau+8/\tau^2}{(6\Theta+8/\tau)^2} \right] 
            \\ &     + 6 \xi \left[ \frac{9\Theta^2+24\Theta/(\eta\tau)+8/(\eta\tau)^2}
                                                   {(6\Theta+8/(\eta\tau))^2}  \right]
         \end{split}      
        \end{equation}
     And the adiabatic index $\Gamma$ is given as ; 
        \begin{equation} 
            \label{eq:CR_EoS_Gamma}
            \Gamma = 1 + \frac{1}{N}
        \end{equation}
     It can be seen that for $\Theta >> 1$ (very high temperature), N approaches 
     asymptotically 3 and $\Gamma \xrightarrow{}4/3$ and for $\Theta << 1$ (very
     low temperature), N approaches asymptotically 3/2 and $\Gamma \xrightarrow{}5/3$.
The CR EoS has been implemented by many authors to study diverse scenarios \citep{2011_Chattopadhyay_Chakrabarti_IJMPD..20.1597C,2014_Cielo_etal_MNRAS.439.2903C,2018_Dihingia_etal_PhRvD..98h3004D,2021joshietal_MNRAS.502.5227J,2022_Joshi_etal_ApJ...933...75J,2023_Sarkar_etal_MNRAS.522.3735S,2023_Joshi_Chattopadhyay_ApJ...948...13J,2024_Debnath_et_al_MNRAS.528.3964D,2024_Joshi_et_al_ApJ...971...13J}.

\section{\label{sec:Numerical Model and Setup}Numerical Model and Setup}
    We numerically solve the hydrodynamic equations \eqref{eq:HD eqns} in two space-dimensions. We solve (a) the classical BHL case as axisymmetric equations in the $(r,z)$ plane and (b) the lateral BHL along the equatorial plane in polar coordinates $(r,\phi)$ utilizing a non-relativistic 
    simulation scheme. We use a second-order Eulerian total variation diminishing (TVD) simulation code \citep{1983_HARTEN_JCoPh..49..357H}. In some parts of our work, we employ the HLLC solver \citep{2009_Toro} of our simulation code. We use an inertial frame of reference that is at rest with respect to the compact object.
  
\subsection{Units and computational grid}  
   We use the geometric units in our analysis where $G = 0.5$ and $c = M = 1$. Therefore, the units of length and time are $\rg = 2GM/c^2$ and $\tg = \rg/c$, respectively, where c is the speed of light. The density is measured in units of the density of the interstellar medium at infinity.
   Our axisymmetric simulations are evolved in a cylindrical domain of size $r \times z = [0:1500] \times [-700:2300]$ in units of the Schwarzschild radius $\rg$, where the domain is discretized into $n_{\rm r}=750$ cells in the r-direction and  $n_{\rm z}=1500$ cells in the z-direction which gives a uniform resolution of $\Delta r = \Delta z =  2$.
   Further, for equatorial plane simulations, we use a non-uniform grid that extends from $\rin=5$ to $\rou=1596$ in the r-direction and covers the whole region from $\phi=0$ to $2\pi$ in $\phi$-direction. The radial domain is logarithmically spaced in 440 cells, and the $\phi$-domain is equispaced in 480 cells for all the cases. The size of the innermost radial cell is $6.59 \times 10^{-2}$, whereas the size of the outermost radial cell is $21.17$. This choice gives $\Delta r / r = \Delta \phi = 1.31 \times 10^{-2}$, thereby maintaining the cell aspect ratio ($\Delta r / r \Delta \phi$) to be constant along a radius.
   Since the calculated accretion radius (Bondi radius, $\rB$) is always less than the domain size, the size of the computational domain is sufficient for all models.

     \begin{ruledtabular}
      \begin{table*}
      	\caption{Details of the parameters used in the simulations}
	      \label{tab:siml_param}
	      \begin{tabular}{lccccccc}
	      	Model & $v_\infty$ & $\mathcal{M}_\infty$ & $\xi$ & $\Theta_\infty$ & $T_\infty [K]$ & $\rB [\rg]$ & Notes \\ 
	      	\tableline
	      	M1p5X1    & 0.05   & 1.5    & 1.0   & $7.49 \times 10^{-4}$   & $4.08 \times 10^9$   & 148 & Axisymmetric/Classical \\
                M1p5Xp5   & 0.05   & 1.5    & 0.5   & $7.57 \times 10^{-4}$   & $2.06 \times 10^9$   & 146 & " \\
                M1p5X0    & 0.05   & 1.5    & 0.0   & $6.67 \times 10^{-4}$   & $3.95 \times 10^6$   & 172 & " \\
                M2X1      & 0.05   & 2.0    & 1.0   & $4.12 \times 10^{-4}$   & $2.24 \times 10^9$   & 210 & " \\
                M2Xp5     & 0.05   & 2.0    & 0.5   & $4.13 \times 10^{-4}$   & $1.12 \times 10^9$   & 210 & " \\
                M2X0      & 0.05   & 2.0    & 0.0   & $3.75 \times 10^{-4}$   & $2.22 \times 10^6$   & 230 & " \\
                M6X1      & 0.05   & 6.0    & 1.0   & $4.25 \times 10^{-5}$   & $2.31 \times 10^8$   & 370 & " \\
                M6Xp5     & 0.05   & 6.0    & 0.5   & $4.23 \times 10^{-5}$   & $1.15 \times 10^8$   & 370 & " \\
                M6X0      & 0.05   & 6.0    & 0.0   & $4.16 \times 10^{-5}$   & $2.47 \times 10^5$   & 370 & " \\
                Mp5X1      & 0.01   & 0.5    & 1.0   & $2.59 \times 10^{-4}$   & $1.41 \times 10^9$   & 668 & Equatorial/Lateral \\
                Mp5Xp5      & 0.01   & 0.5    & 0.5   & $2.58 \times 10^{-4}$   & $7.03 \times 10^8$   & 678 & " \\
                Mp5Xp2      & 0.01   & 0.5    & 0.2   & $2.50 \times 10^{-4}$   & $2.73 \times 10^8$   & 718 & " \\
	      \end{tabular}
            \tablenotetext{}{\textbf{Note.} \justifying The variables $v_\infty$ and $\mathcal{M}_\infty$ represent the velocity and the Mach number of injected flow at infinity. $\xi$, $\Theta_\infty$, $T_\infty$ represent the composition parameter, dimensionless temperature, and physical temperature (in K) of the flow at infinity, respectively. $\rB$ represents the Bondi radius calculated analytically from the energy balance in units of the Schwarzschild radius $\rg$.}
      \end{table*}       
   \end{ruledtabular}

\subsection{Initial setup and Boundary conditions}
  Classical BHL: The density and the velocity are taken to be uniform and constant in the computational region. Matter is continuously injected with density $\rho_\infty$, pressure $p_\infty$ and velocity $v_\infty$ into the domain from the upstream boundary ($z=Z_{\rm min}$) in the axisymmetric simulation. The injection Mach number is $\mathcal{M}_\infty=v_\infty/c_{s\infty}$ and $c^2_{s\infty}=\Gamma_\infty p_\infty/\rho_\infty$.  
  The outflow boundary condition is imposed on the outer r-boundary ($R_{\rm max}$) and the downstream z-boundary ($z=Z_{\rm max}$). 
  Therefore the outer r and z boundaries are causally disconnected from the accretion flow \citep{2014_Lee_ApJ...783...50L, 2015_mellah_casse_MNRAS.454.2657E, 2019_xu_stone_MNRAS.488.5162X}. Further, the inner boundary ($r \leq r_{\rm in} = 6\rg$) is modeled as an absorbing boundary or sink so that no physical information can propagate from inside the boundary to the outside \citep{1994a_Ruffert_A&AS..106..505R}. 
  The size of the accretor is smaller than the size of the sink. 
 
   Lateral BHL: In equatorial plane simulations, for a continuous parallel wind at large distances ($v^2=v_\infty^2$), the initial velocity field components of the injected gas are specified in terms of an asymptotic velocity $v_\infty$ ($v_{\rm r} = v_\infty cos \theta$,  $v_{\rm \theta} = -v_\infty sin \theta$).
   Further, the outer boundary ($r=R_{\rm max}$) is split into two halves: one portion extending from 265 to 270 degrees in which the gas enters the domain where we apply inflow boundary conditions and a second portion where the gas leaves the domain and we apply outflow boundary conditions. In addition, in the angular domain, we use periodic boundary conditions.

  We have performed simulations for various values of composition parameters $\xi$ and Mach number $\mathcal{M}_\infty$, which are listed in Table~\ref{tab:siml_param}. The first column is the model names, the second and the third columns represent the injection velocity and Mach number of the flow, and the fourth column is the composition of the flow. The dimensionless temperature ($\Theta_\infty$) at the injection radius is in column 5, the corresponding temperature is in column 6, and the Bondi radius is listed in the seventh column. In order to numerically compute the mass accretion rate, we use a detector evaluating the local mass accretion rate ($\rho v r^2$) at different radii and then add them up for the total mass accretion rate at each time step.

\section{\label{sec:Results}Results}
\subsection{Classical BHL: Axisymmetric simulations}
\subsubsection{Morphology of the flow for different models} 
   \begin{figure*}
       \includegraphics[width=\textwidth]{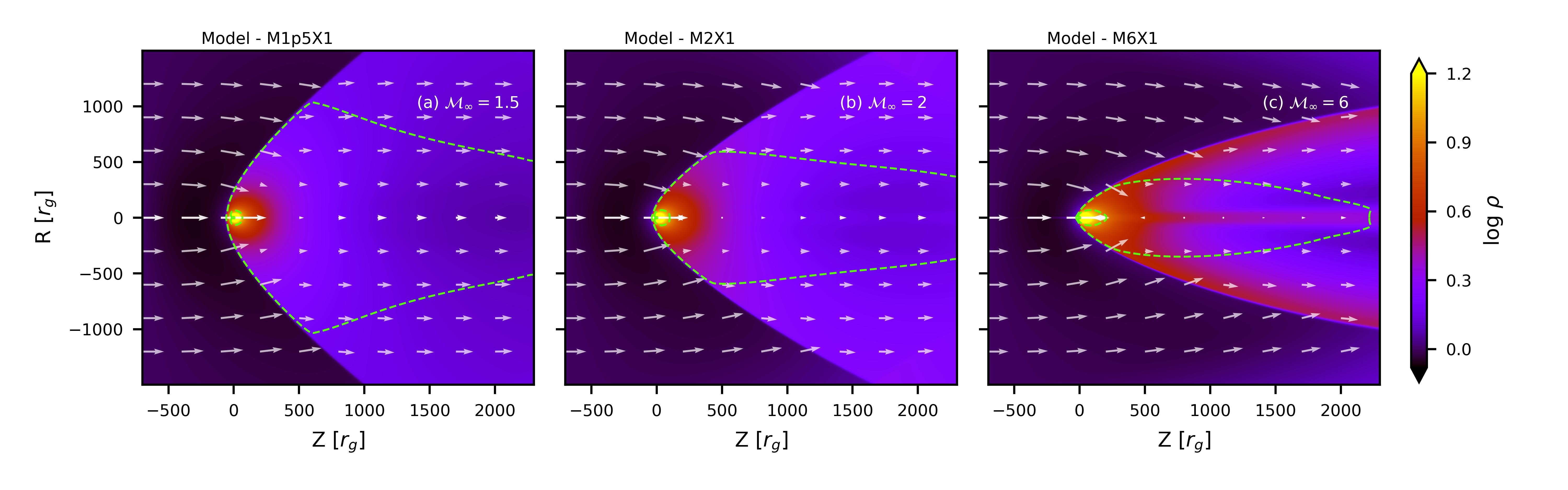}
       \caption{Hydrodynamic simulations of the BHL accretion to a compact object moving supersonically through an ambient medium. Logarithmic color maps (arbitrary units) of the stationary density profiles for various models having Mach numbers $\mathcal{M}_\infty$ = 1.5, 2, and 6 in panels (a), (b), (c) respectively along with the velocity vectors (white) overplotted at time t=10.0 $\times$ $10^4$ $\rg/c$. The length of the arrows is proportional to the velocity. The green dashed line shows the Mach-1 surface in the flow. The length is measured in units of the Schwarzschild radius $\rg$.}
       \label{fig:density_morphology}
   \end{figure*}

   \begin{figure*}
       \includegraphics[width=\textwidth]{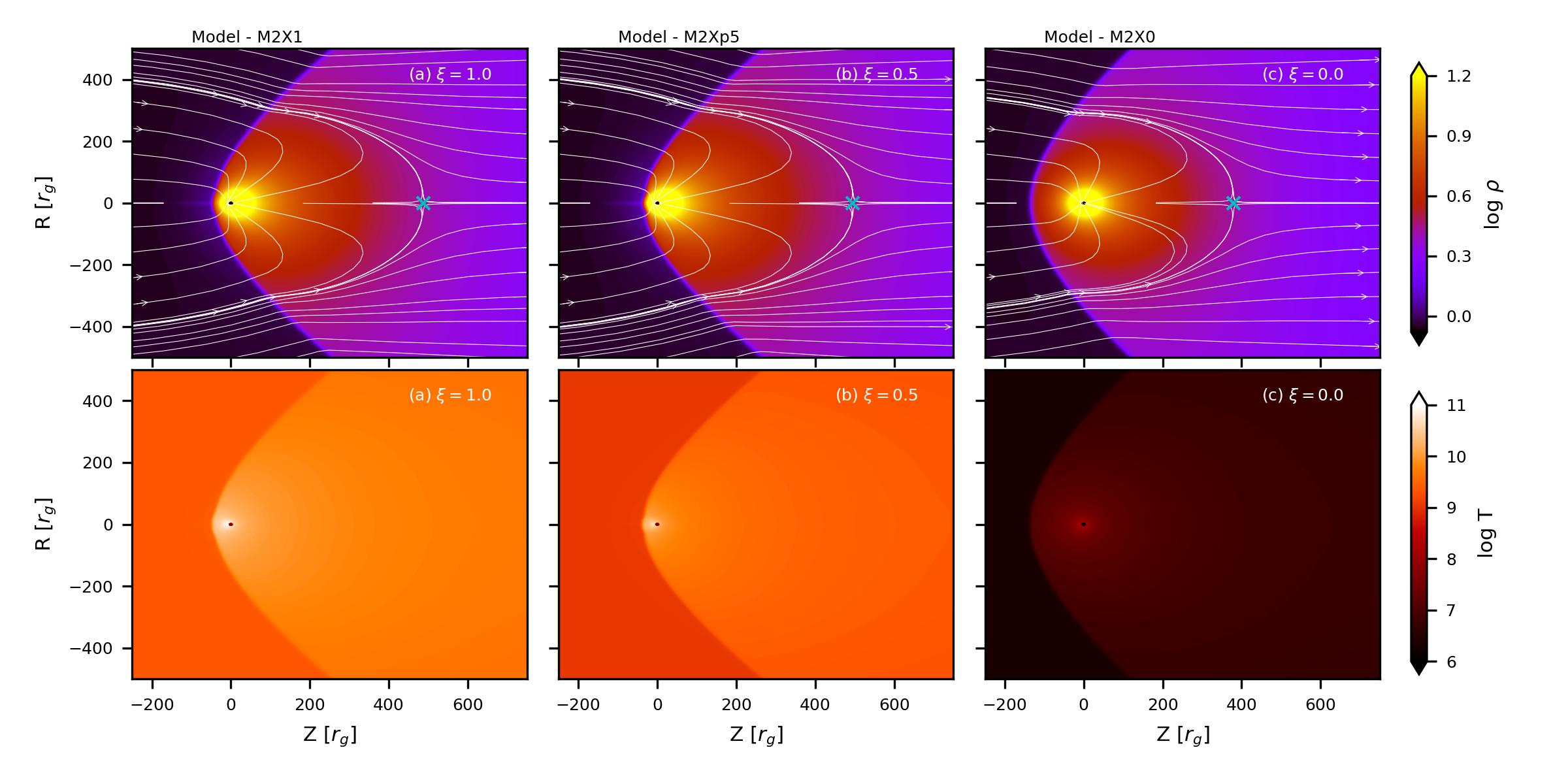}
       \caption{BHL accretion flows to a compact object moving with Mach number $\mathcal{M}_\infty=2$ at late time t=20.0 $\times$ $10^4$ $\rg/c$, when the flow is in the quasi-steady state. The background color represents the density in the upper and the temperature in the lower panels. The flow has different compositions $\xi$ = 1.0, 0.5, and 0.0 in panels (a), (b), and (c), respectively. The solid lines in white color represent streamlines with arrows showing the direction of the velocity. A cyan cross is used to mark the position of the stagnation point.}
       \label{fig:composition_effect}
   \end{figure*}
   
 As the simulation is turned on, the supersonic flow forms a bow shock encircling the central acceretor from three sides, with increased density, temperature, and lower velocity in the post-shock region. The shock front is initially narrow, but then it spreads in a wider angle till it reaches a steady state.  
 Figure~\ref{fig:density_morphology} shows the morphology of the BHL accretion flow for models M1p5X1, M2X1, and M6X1, respectively. In all three models, the relative velocity between the accretor and the flow is supersonic, and the composition parameter is kept the same ($\xi=1.0$). The material coming from the upstream boundary ($z=Z_{\rm min}$) would be gravitationally attracted, and it would be compressed as it approaches the accretor. The internal pressure of the flow will increase because of the compression, causing significant disruption and the formation of a bow shock. The shock front developed around the accretor is known as the Mach cone. The opening angle of this Mach cone will depend on the Mach number of the flow as $\theta=\sin^{-1}(1/\mathcal{M})$. The more the supersonic flow is, the more it would be confined in that cone, which can be seen in Figure~\ref{fig:density_morphology} (a), (b), and (c) having different Mach numbers ($\mathcal{M}_\infty$).
 To investigate the effect of the composition on the accretion flow, we examine models having the same Mach number ($\mathcal{M}_\infty$) and different composition parameters $\xi$. Figure~\ref{fig:composition_effect} shows the accretion flow for models M2X1 (i.e., $\xi=1.0$), M2Xp5 ($\xi=0.5$), and M2X0 ($\xi=0.0$) in panels (a), (b), and (c), respectively. It is actually a flow with injection Mach 2, but different compositions fall onto the accretor. The top panels (a, b, \& c) show the density contours of the three models; overlaid are the streamline of the flow (white lines). The bottom panels (a, b, \& c) show the contours of temperature.
 Both are in log scale. The streamlines having impact parameters less than the Bondi radius $\rB$ are captured by the accretor, and those having impact parameters larger than the Bondi radius escape outward, as seen in Figure~\ref{fig:composition_effect}. The distance between the blackhole (located at the origin) and the stagnation point (marked as a cyan cross) on the z-axis is the measure of the impact parameter of the flow and is slightly more than the $\rB$. The position of the shock front is slightly different for flows having different composition parameters $\xi$. The shock front is farther away from the accretor for electron-positron flow, or $\xi=0.0$, compared to the case of electron-proton flow $\xi=1.0$ or the flow with an equal proportion of positrons and protons i. e., $\xi=0.5$. 
 
   \begin{figure*}
   \begin{center}
      \includegraphics[width=15cm]{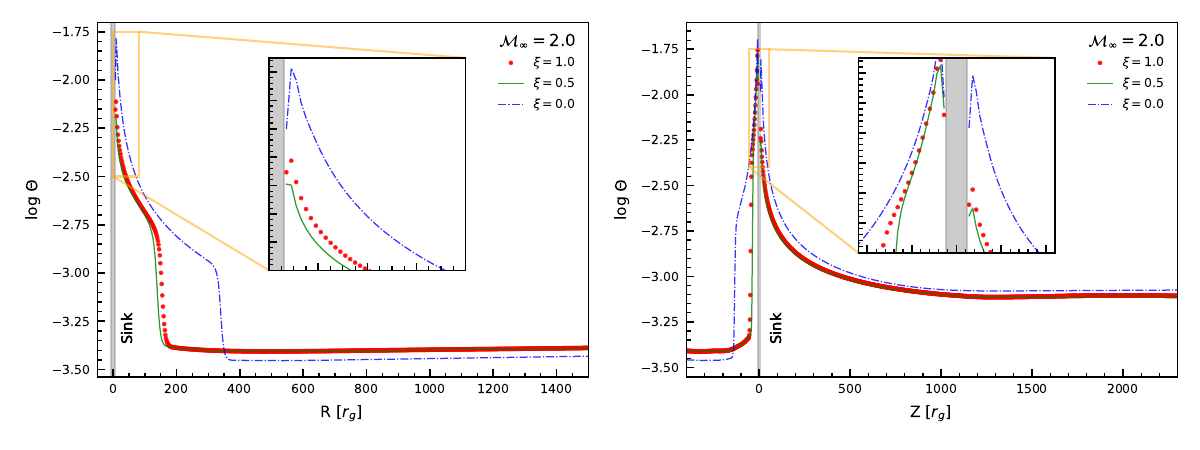}
   \end{center}
       \caption{Radial and axial variation of dimensionless temperature ($\Theta$) of the accretion flow is plotted in left and right panels, respectively, for $\mathcal{M}_\infty$ = 2. Different compositions $\xi$ = 1.0, 0.5, and 0.0 are shown by the red dotted line, green solid line, and blue dash-dotted line, respectively. The shaded region in the grey color is the sink region around the central object.}
       \label{fig:spline_plot_THETA}
   \end{figure*}

   \begin{figure*}
    \includegraphics[width=\textwidth]{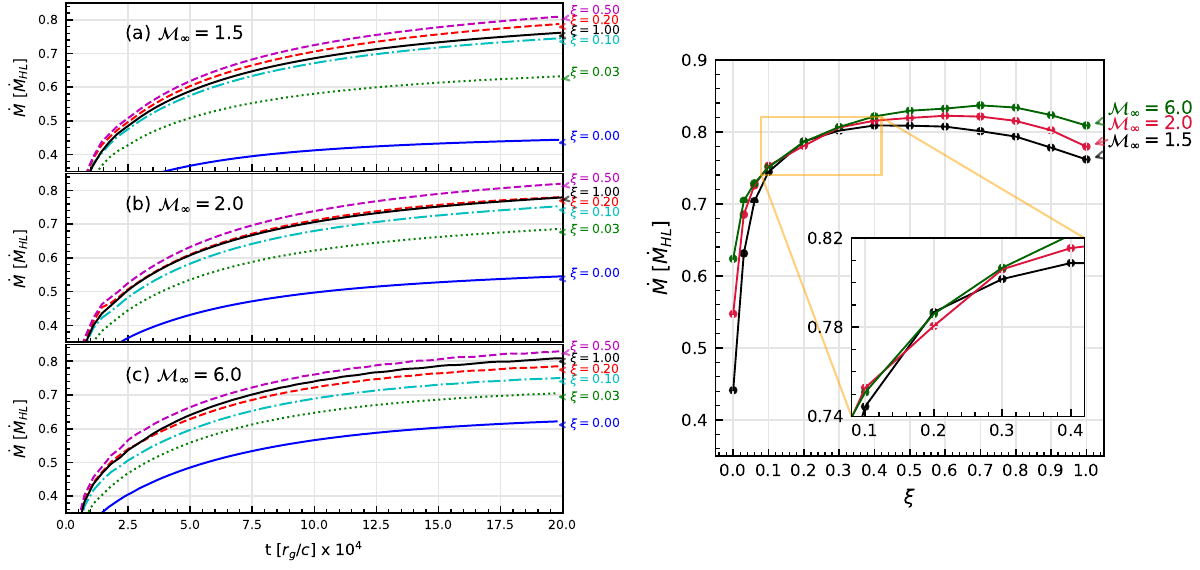}
       \caption{Left: Time evolution of the numerically calculated mass accretion rates for different Mach numbers $\mathcal{M}_\infty$ = 1.5, 2, and 6 are shown in panels (a), (b), and (c), respectively. Different compositions $\xi$ = 0.0, 0.03, 0.1, 0.2, 0.5, 1.0 are shown in blue, green, cyan, red, purple, and black colors, respectively. Right: Mass accretion rates for all models with different $\mathcal{M}_\infty$ and $\xi$ at late time t=20.0 $\times$ $10^4$ $\rg/c$ are shown when the flow is in the quasi-steady state. The time is measured in units of $\rg/c$ and the mass accretion rate $\dot{M}$ is normalized to the Hoyle–Lyttleton accretion rate $\dot{M}_{\rm HL}$ given by equation \eqref{eq: HL accretion rate}.}
       \label{fig:accretion_rates}
   \end{figure*}
   
 Figure~\ref{fig:spline_plot_THETA} shows the radial and axial variation of $\Theta$ of the flow for different composition models M2X1, M2Xp5, and M2X0. $\Theta$ represents the ratio of the thermal energy to the rest mass energy of the flow. As one approaches the accretor along the $r$ direction or $\pm z$ directions, $\Theta$ increases strongly because of the adiabatic compression. 
 The post-shock region for the {electron-positron} flow ($\xi=0.0$) is wider. {In most of the regions for electron-positron flow ($\xi$ = 0.0), the adiabatic index ($\Gamma$) is approximately 1.67, therefore thermally non-relativistic. In other cases with different $\xi$ values, the adiabatic index ($\Gamma$) is typically around 1.5, decreasing to about 1.4 near the accretor. Therefore, flows with protons are hotter. This indicates that electron-positron flow is colder compared to other cases, and the thermal forces are insufficient to provide the barrier for the incoming supersonic flow, to form a shock. Therefore, it requires a larger volume of matter to do that resulting in a wider post-shock region.}

\subsubsection{Effect of the composition at mass accretion rates}
 Figure~\ref{fig:accretion_rates} presents the evolution of the numerically computed mass accretion rates in the vicinity of the inner boundary as a function of time for various injected Mach numbers and composition parameters. The mass accretion rate $\dot{M}$ is normalized to the Hoyle–Lyttleton accretion rate $\dot{M}_{\rm HL}$ given by equation \eqref{eq: HL accretion rate}, which depends on the density and the velocity of the flow at infinity. It may be noted that the sound speed at the injection is controlled by tuning the Mach number at infinity and the composition. $\dot{M}_{\rm HL}$ doesn't depend on the sound speed or the temperature of the flow, therefore, the normalization is the same for all of our models. Mass accretion rates in every case saturate to a value lower than $\dot{M}_{\rm HL}$. Since we are injecting with the same velocity, the higher Mach number indicates lower temperature, which results in larger $\rB$ for the flow. And higher $\rB$ implies higher mass accretion rate. Further, an increase in Mach number beyond 6 results in a marginal increase in accretion rate, which would saturate because of the dominant kinematical effects rather than the thermal. Moreover, in the case of models with identical $\mathcal{M}_\infty$, the mass accretion rate appears to be decreased when considering purely leptonic flows ($\xi = 0.0$) compared to the other cases. However, when comparing electron-proton flow ($\xi = 1.0$) to mixed flow ($\xi = 0.2$) for $\mathcal{M}_\infty = 1.5$, the mass accretion rate is lower in the former, while the opposite is true for $\mathcal{M}_\infty = 6.0$. While for $\mathcal{M}_\infty = 2.0$, flows of both the compositions $\xi = 1.0, ~0.2$ are having approximately the same mass accretion rates.
 Further, the right panel of figure~\ref{fig:accretion_rates} shows the mass accretion rates for all the models with different $\mathcal{M}_\infty$ and $\xi$ at late time t=20.0 $\times$ $10^4$ $\rg/c$ when the flow has achieved the quasi-steady state. For the mixed flow between $\xi$ = 0.5 to $\xi$ = 0.7, accretion is peaking to a higher value, while it is lower for the electron-proton flow or purely electron-positron flow.
 {It may be noted that the flow is thermally relativistic if its thermal energy (average kinetic energy of the constituent particles of the gas) is comparable to or higher than its inertia. An electron-proton gas has a high temperature because the average momentum transferred by its particles is higher, but by the same token also has high inertia. An electron-positron pair plasma, on the other hand, has the least inertia, but the average momentum transferred between the constituent particles is also very low. The competition between inertia and thermal energy peaks for intermediate values of $\xi$ and hence are thermally more relativistic than either electron-proton or electron-positron flow. The accretion rate for classical BHL is maximized when the incoming supersonic flow pattern is stalled and diverted toward the central black hole. Flows with intermediate $\xi$ being relatively hotter, offer a perfect barrier to divert the supersonic matter into the black hole. Therefore the accretion rate peaks for intermediate values of $\xi$.}

   \begin{figure*}
       \includegraphics[width=\textwidth]{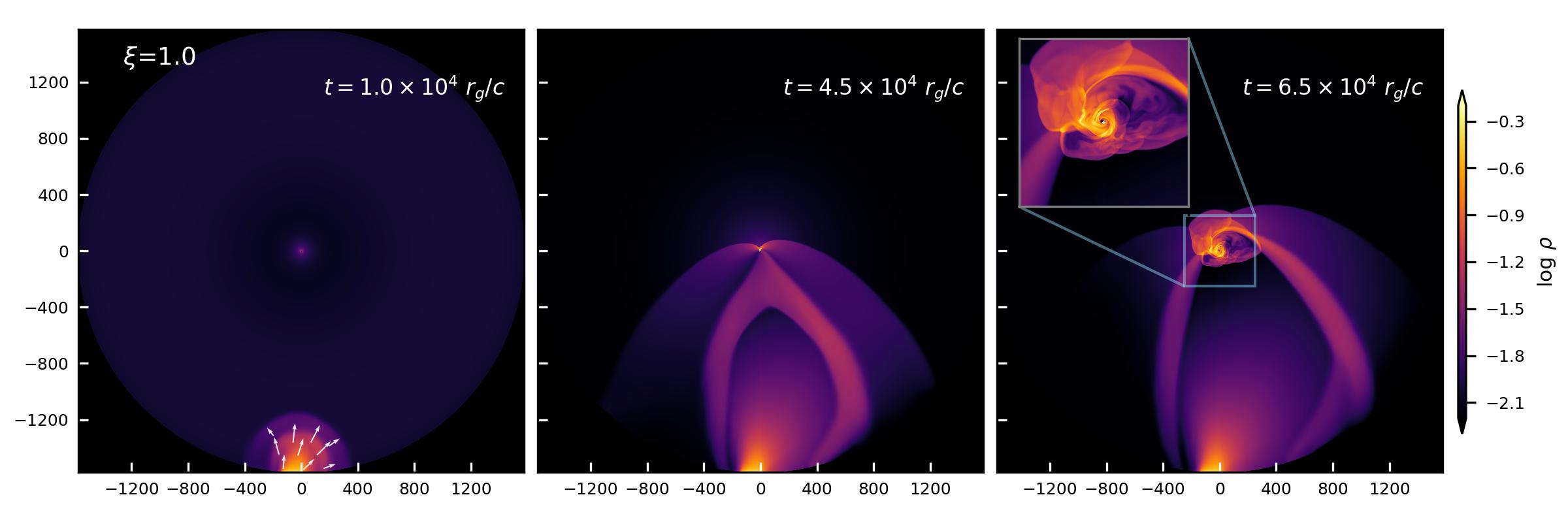}
       \caption{{Contours of logarithmic density plotted at various times t = 1.0, 4.5, 6.5 $\times$ $10^4$ $\rg/c$ in left, middle, right panels respectively for Model Mp5X1 ($\xi=1.0$). The white arrows in the left panel are the velocity vectors of the injected flow. The inset panel shows the zoomed region of the accretion flow. The length is measured in units of the Schwarzschild radius $\rg$.}}
       \label{fig:Eqt_injectionframes}
   \end{figure*}

   \begin{figure*}
       \includegraphics[width=\textwidth]{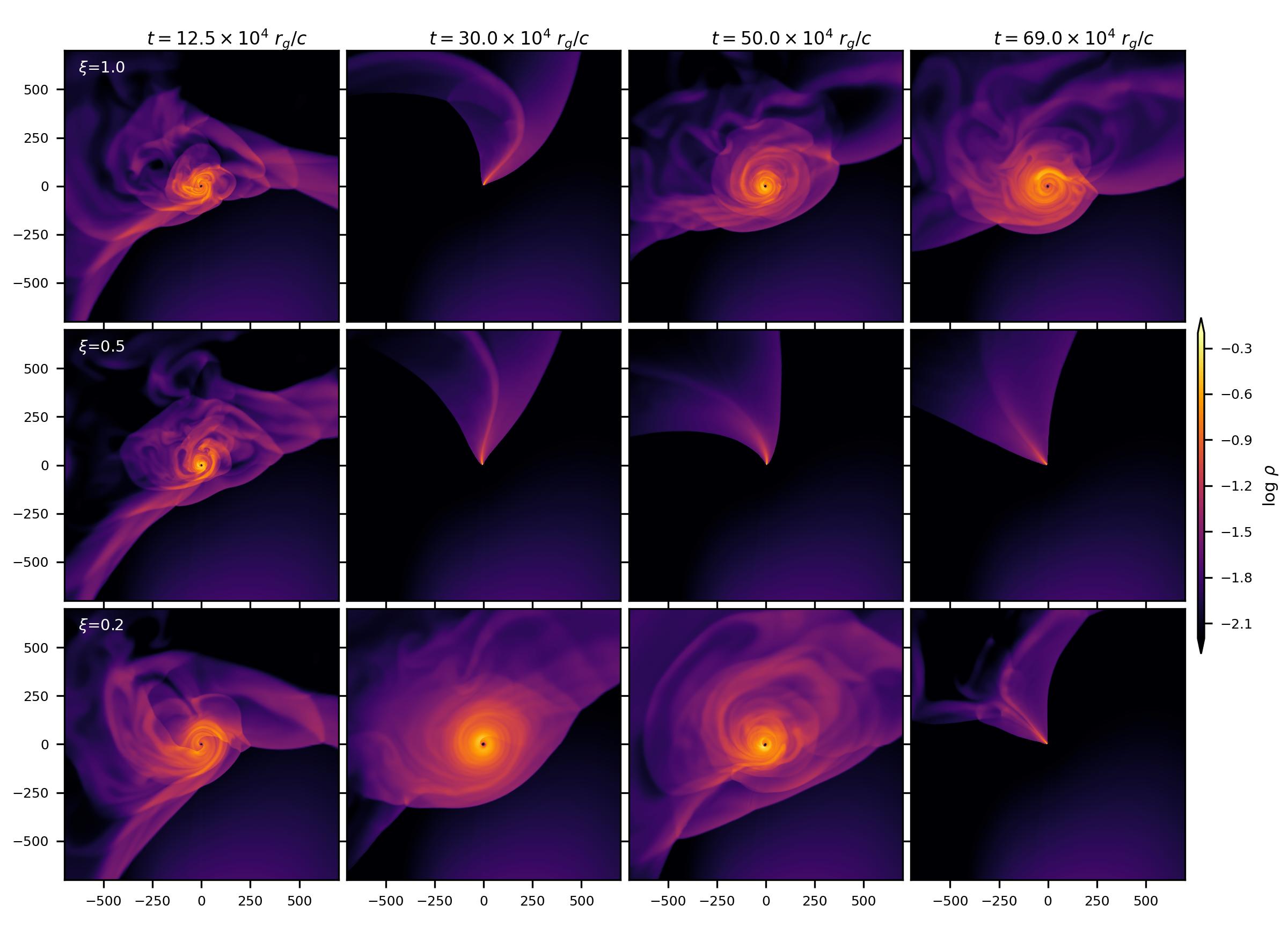}
       \caption{Snapshots of logarithmic density distribution at different times t = 12.5, 30.0, 50.0, 69.0 $\times$ $10^4$ $\rg/c$ are shown in different columns for models Mp5X1 ($\xi=1.0$), Mp5Xp5 ($\xi=0.5$), and Mp5Xp2 ($\xi=0.2$) in upper, middle, and lower panels, respectively. The length is measured in units of the Schwarzschild radius $\rg$. }
       \label{fig:Eqt_density_morphology}
   \end{figure*}
   
\subsection{Lateral BHL: Equatorial simulations}
\begin{figure*}
       \includegraphics[width=\textwidth]{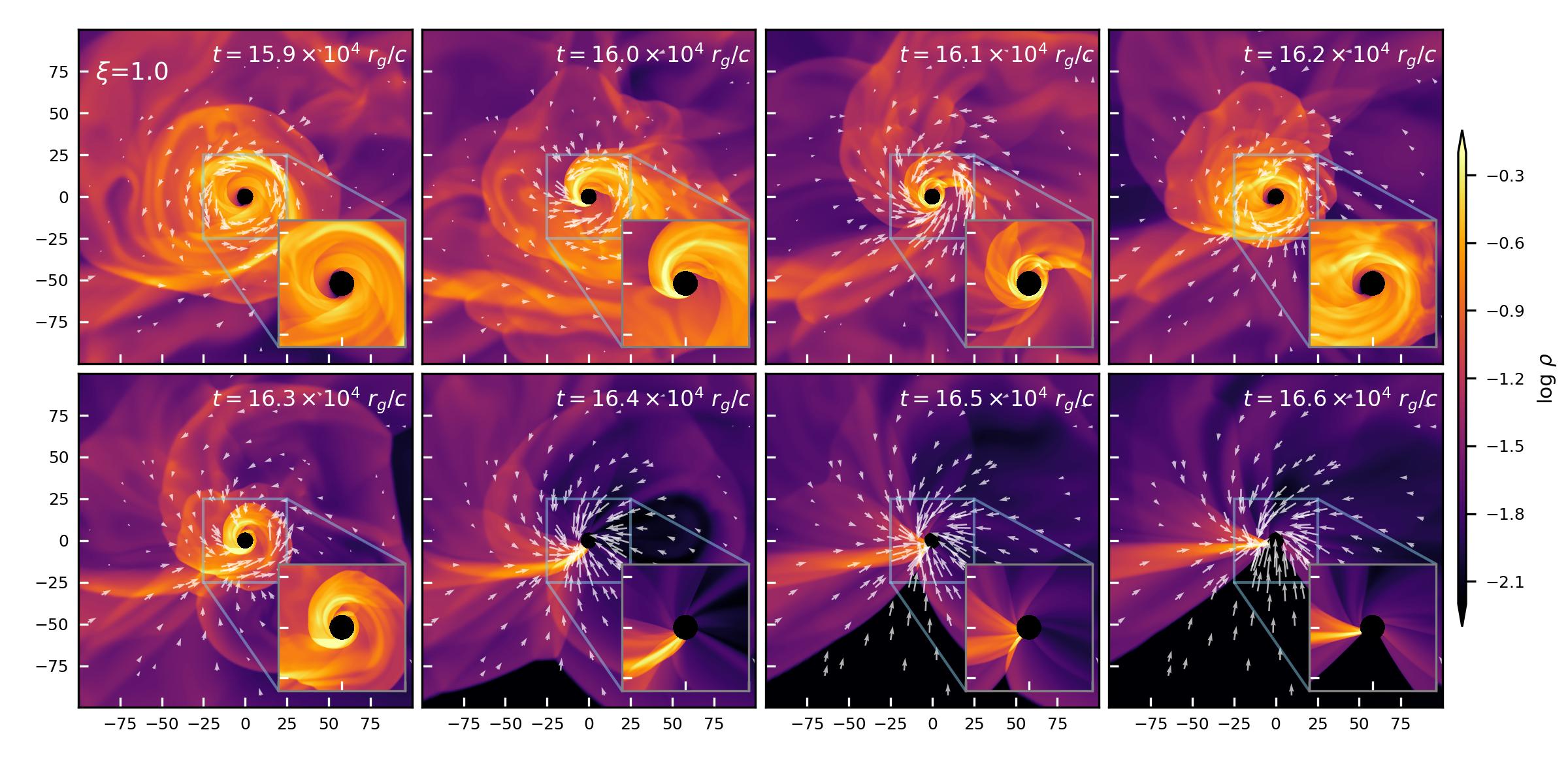}
       \caption{{Contours of logarithmic density plotted at various times t = 15.9---16.6 $\times$ $10^4$ $\rg/c$ in different panels for Model Mp5X1 ($\xi=1.0$) showing the transition of the disk into a shock cone. The inset panel shows the zoomed region near the accretor. The arrows in the white are the velocity vectors of the flow.} }
       \label{fig:Eqt_disky_cone_transition}
   \end{figure*}

\begin{figure*}
       \includegraphics[width=\textwidth]{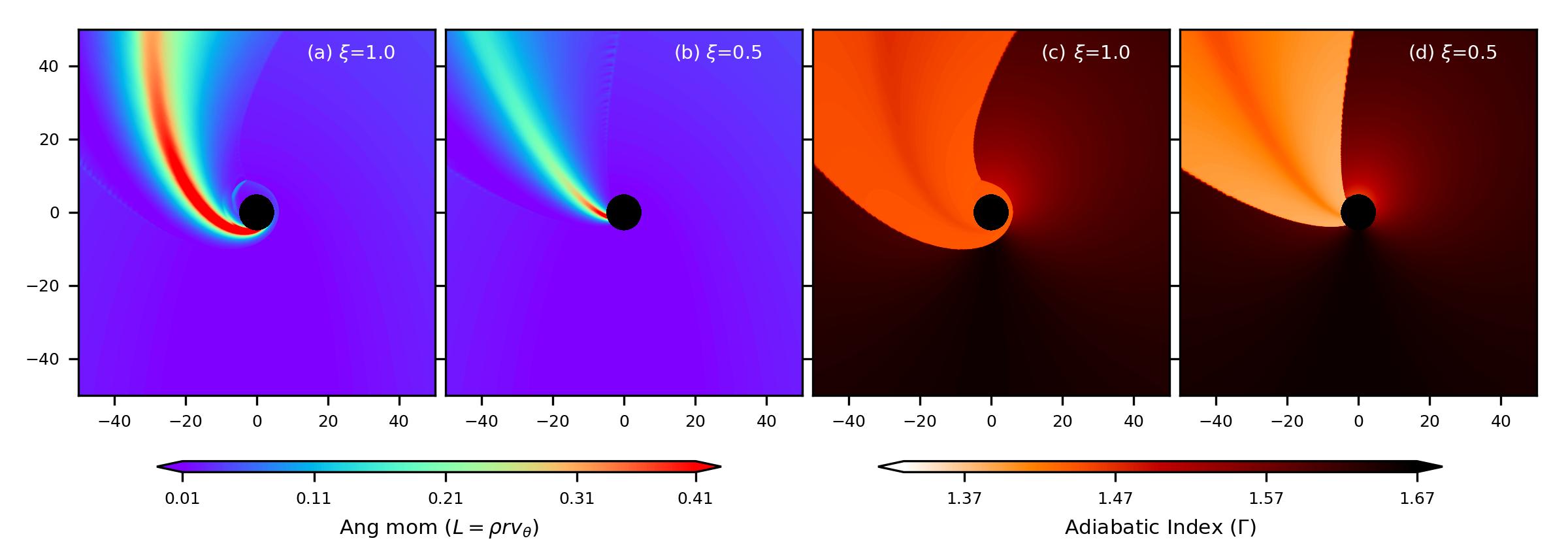}
       \caption{Contours of angular momentum density (a $\&$ b) and adiabatic index $\Gamma$ (c $\&$ d) are shown in the inner region of accretion flow for model Mp5X1 (in panels a $\&$ c: $\xi$=1.0) and Mp5Xp5 (in panels b \& d: $\xi=0.5$) at t=33.5 $\times$ $10^4$ $\rg/c$ just before disk reformation.} 
       \label{fig:Eqt_angmom_gamma}
   \end{figure*}

\begin{figure*}
       \includegraphics[width=\textwidth]{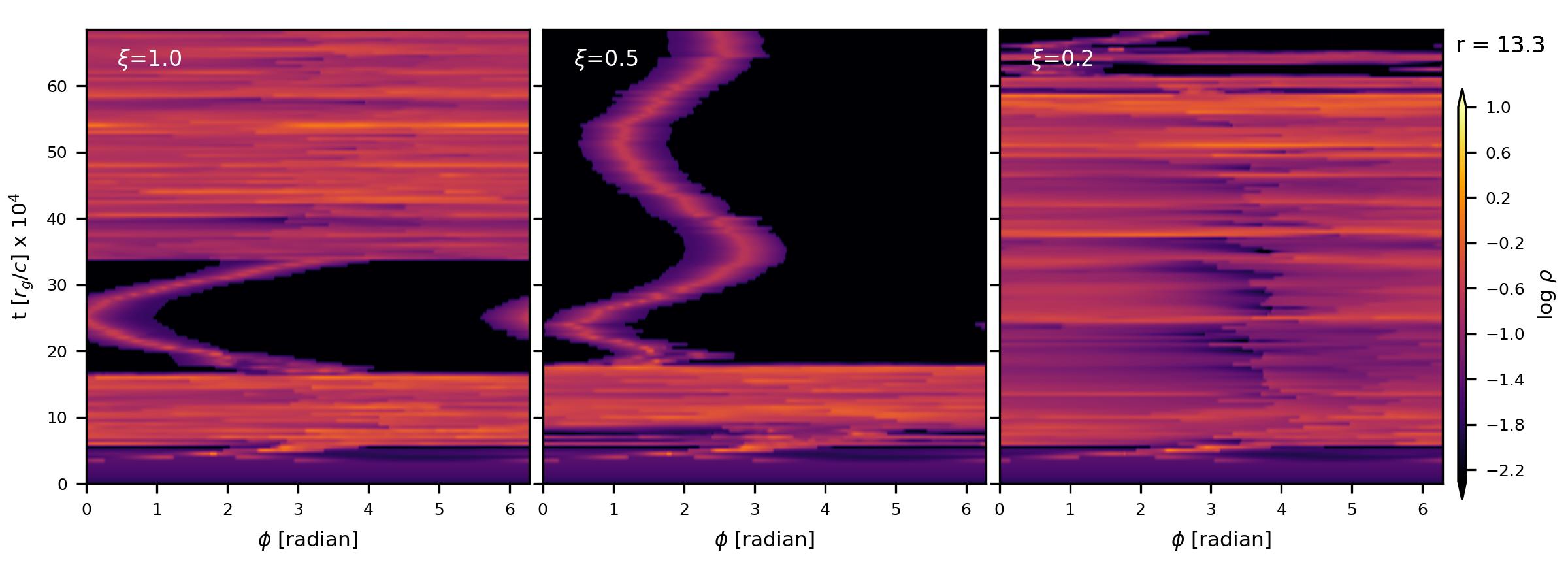}
       \caption{The time-dependent variation of logarithmic density as a function of $\phi$ at fixed radius $r=13.3\rg$ is shown. The left panel is for model Mp5X1 ($\xi=1.0$), the middle one for Mp5Xp5 ($\xi$=0.5), and the right panel for Mp5Xp2 ($\xi$=0.2). }
       \label{fig:Eqt_t_phi_frames}
   \end{figure*}

  {Figure~\ref{fig:Eqt_injectionframes} shows the early-stage evolution of our lateral BHL simulation for Model Mp5X1 ($\xi=1.0$). As the simulation is turned on, the wind coming from the inflow boundary encircles the black hole and forms a disk-like structure (right panel of Figure~\ref{fig:Eqt_injectionframes}) with higher density in the central region.}
  {In Figure~\ref{fig:Eqt_density_morphology}, we compare} the contours of the logarithmic density distribution of accretion flow for three models Mp5X1 ($\xi=1.0$), Mp5Xp5 ($\xi=0.5$), and Mp5Xp2 ($\xi=0.2$), respectively having different composition parameters in three-row panels. 
  At the beginning of the simulation, {one can see a disk structure for all three models}
  in the first column of Figure~\ref{fig:Eqt_density_morphology} at t=12.5 $\times$ $10^4~\rg/c$. 
  As the simulation progresses, 
  this disk transforms into a shock cone temporarily (see the second column of Figure~\ref{fig:Eqt_density_morphology} for $\xi$=1.0, 0.5) but reappears for $\xi=1$ at a later time. 
  {Figure~\ref{fig:Eqt_disky_cone_transition} presents the density contours and the velocity vectors (arrows) at 8-time snaps with a cadence of $\Delta t=0.1 \times 10^4~\rg/c$ highlighting the transition of the disk into a shock cone for Model Mp5X1 ($\xi=1.0$). 
  One can see that the disk formed is not axis-symmetric; spiral structures form at some time, and at another, those structures circularize. Therefore the balance of forces around the accretor is not uniform, which causes uneven accretion around the accretor. This may disrupt the spiral disk structure of the flow, increasing the accretion rate and formation of the shock cone. This is also seen in the panels at $t=16.4\times 10^4~\rg/c$, where the velocity vectors are dominant and into the accretor in the region where accretion has occurred.}
  Once the shock cone forms, it is subjected to the flip-flop instability \citep{1987_Matsuda_MNRAS.226..785M,1988_Fryxell_Taam_ApJ...335..862F,2011_Donmez_MNRAS.412.1659D}, where the whole shock cone oscillates due to the tangential velocities in the system, but oscillations are more pronounced for $\xi=1$ compared to $\xi$=0.5 as can be seen in the contours of angular momentum density distribution (Figure~\ref{fig:Eqt_angmom_gamma}a \& b) for both $\xi=1.0~\&~0.5$ at t=33.5 $\times$ $10^4$ $\rg/c$. Soon after, the disk reappears for $\xi=1.0$ but not for $\xi=0.5$. {The angular momentum density is defined as $L=\rho r v_\theta$.}
  We plot the angular momentum density contours of the shock cone around the BH for $\xi=1$ (Figure \ref{fig:Eqt_angmom_gamma} a) and for $\xi=0.5$ (Figure \ref{fig:Eqt_angmom_gamma} b), and show that the $\xi=1$ matter has much higher angular momentum {(${L \sim 0.4}$) than $\xi=0.5$ matter has (${L \sim 0.2}$).}
  Since the lower value of $\Gamma$ or the adiabatic index represents higher heat content in the flow, than a higher $\Gamma$ value,
  we also plot the corresponding $\Gamma$ in panels Figure \ref{fig:Eqt_angmom_gamma} c \& d, which shows that the thermal content of $\xi=0.5$ is higher ($\Gamma \sim 1.4$) than $\xi=1.0$ ($\Gamma \sim 1.45$). This implies that the increased random motion of the particles for $\xi=0.5$ hinders the ordered rotational motion. 
  So, for $\xi$=0.5, the disk vanishes and does not reappear because of insufficient angular momentum within the shock cone. For $\xi$=0.2, the disk structure persists for a significantly longer duration. 
  To gain a better understanding of the disk structure and shock cone around the black hole, we present in Figure~\ref{fig:Eqt_t_phi_frames}, the time-dependent variation of logarithmic density as a function of azimuthal direction from the initial time up to the end of the simulation. This figure involves plotting the density as a function of $\phi$ and time $t$ at r = 13.3$\rg$, thereby revealing the temporal behavior of the system at this fixed radius.
  For $\xi$=1.0, the disk temporarily transforms into a shock cone, where densities are confined to a certain angular region (left panel of Figure~\ref{fig:Eqt_t_phi_frames}); the shock cone undergoes flip-flop instability before the disk reappears and persists until the end of the simulation. In contrast, for $\xi$=0.5, the shock cone oscillates, but the disk does not re-appear (middle panel of Figure~\ref{fig:Eqt_t_phi_frames}). For $\xi$=0.2, the disk structure remains intact for a significantly longer duration (right panel of Figure~\ref{fig:Eqt_t_phi_frames}). If the proton proportion is reduced even further i.e., as $\xi \rightarrow 0$, the disk forms and persists.
   
\begin{figure*}
       \includegraphics[width=\textwidth]{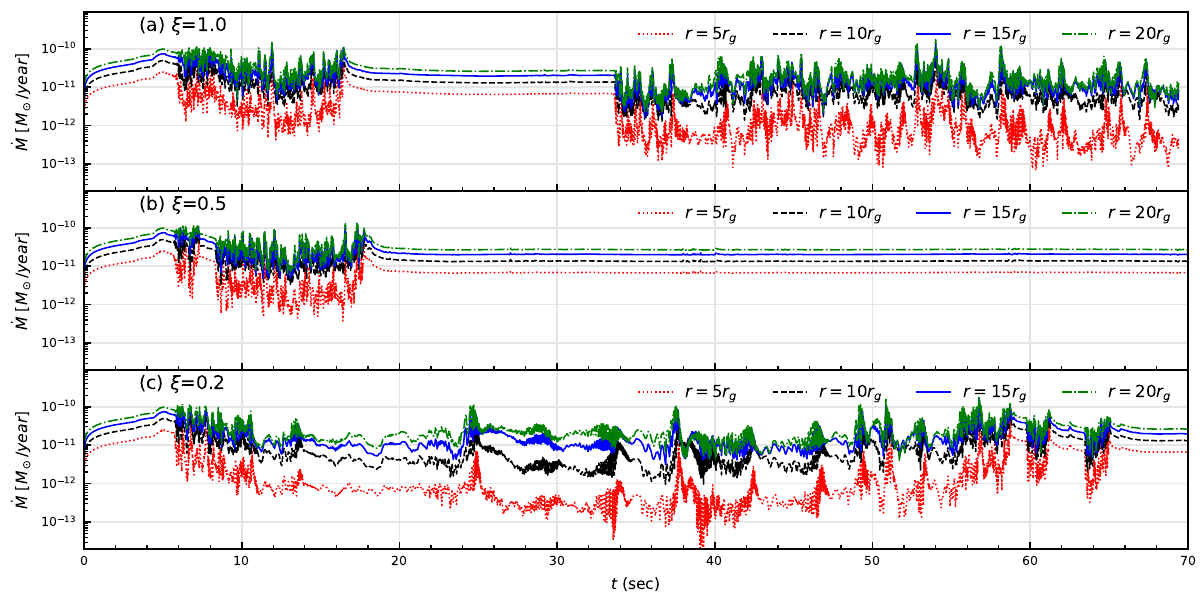}
       \caption{Time evolution of the numerically calculated mass accretion rates for models Mp5X1 ($\xi=1.0$), Mp5Xp5 ($\xi=0.5$), and Mp5Xp2 ($\xi=0.2$) are shown in panels (a), (b), and (c), respectively. Different colors show accretion rates at different radii marked on the top of the panels. The mass of the black hole ($M_{BH}$) is taken to be $10M_\odot$, and the normalized density of the injected wind ($\rho_{\infty}$) is $10^{-11} $ gm/cc. }
       \label{fig:Eqt_accretion_rates}
   \end{figure*}

 Figure~\ref{fig:Eqt_accretion_rates} presents the temporal evolution of the mass accretion rates for three cases of $\xi$ in different panels. We calculate the mass accretion rates at different radii r = 5$\rg$, 10$\rg$, 15$\rg$, and 20$\rg$ shown in red dotted, black dashed, blue solid, and green dash-dotted, respectively. The time is normalized by taking the mass of the black hole to be $10M_\odot$, and the accretion rate is normalized by taking the density of the injected wind to be $10^{-11} $ gm/cc. As can be seen, the accretion rate rises initially and fluctuates up to two orders of magnitude in the disk phase at each radius but approaches an asymptotic value in the shock cone phase.
 {The transition from the disk to cone phase is marked by a higher accretion rate, and the cone to disk phase by the opposite trend.}

 We compute the posteriori radiation emitted by the flow by considering bremssstrahlung and synchtrotron emission. In order to calculate emission due to electron-ion and electron-electron bremsstrahlung, we use the bremsstrahlung cooling rate per unit volume (in ergs/cc/sec) as prescribed in \citet{1982_Svensson_ApJ...258..335S,1983_Stepney_Guilbert_MNRAS.204.1269S,1995_Narayan_Yi_ApJ...452..710N} as,
 
     \begin{equation}
        \label{eq:Qbr_NY}
        Q_{br} = Q_{ei} + Q_{ee},
       \end{equation}

      \begin{eqnarray}
          \label{eq:}
          Q_{ei} = 1.48 \times 10^{-22} {n_{\rm e}}^2 \times 
                            \begin{cases}
                                4 \left(\frac{2\Thetae}{\pi^3}\right)^{0.5} (1 + 1.781\Thetae^{1.34}), & \text{if $\Thetae < 1$},\\
                                \frac{9 \Thetae}{2\pi} \left[ \text{ln}(1.123\Thetae + 0.48) + 1.5 \right], & \text{if $\Thetae > 1$}.
                            \end{cases}  \\  
          Q_{ee} = \begin{cases}
                        2.56 \times 10^{-22} {n_{\rm e}}^2 \Thetae^{1.5} (1 + 1.1\Thetae + \Thetae^{2} -1.25\Thetae^{2.5}), & \text{if $\Thetae < 1$},\\
                        3.40 \times 10^{-22} {n_{\rm e}}^2 \Thetae (\text{ln}(1.123\Thetae) + 1.28), & \text{if $\Thetae > 1$}.
                   \end{cases}
      \end{eqnarray}

 where $n_{\rm e}$ is the number density of electrons, and $\Thetae=kT/ \mel c^2$ is the dimensionless electron temperature. Further, we also consider the contributions from thermal synchrotron emissivity in the presence of a stochastic magnetic field. The synchrotron cooling rate is given by following \citet{1983_Shapiro_Teukolsky_bhwd.book.....S} as,

      \begin{equation}
          \label{eq:Qbr_ST}
          Q_{syn} = \frac{16}{3} \frac{e^2}{c} \left(\frac{eB}{\mel c}\right)^2 \Thetae^2 n_{\rm e}
      \end{equation}

 where $B$ is the stochastic magnetic field, which is estimated by assuming the magnetic pressure is $1\%$ of the thermal gas pressure. To estimate the electron temperature, we assume the adiabatic index, $\Gamma$, is the same for ions and electrons, i.e., $\Gamma_{\rm i} = \Gamma_{\rm e} = \Gammaeq$. Following this, if equation \eqref{eq:CR_EoS_Gamma} is modified for only electrons, one may get a quadratic equation, $\Thetae^2 (36 - 27\Gammaeq) + \Thetae (48-36\Gammaeq) + (10-6\Gammaeq) = 0$, which can be solved analytically for $\Thetae$, provided $\Gammaeq$ for any value of $\xi$ \citep{2014_Kumar_Chattopadhyay_MNRAS.443.3444K}. 
 Integrating the expressions \eqref{eq:Qbr_NY}, \eqref{eq:Qbr_ST} over the whole volume by summing up all the contributions from each point in the domain, we get the bolometric luminosity of the system, ($L_{\rm bol}=\int (Q_{br}+Q_{syn})dV$). The emission from outside the simulation domain is neglected in all calculations.

   \begin{figure*}
       \includegraphics[width=\textwidth]{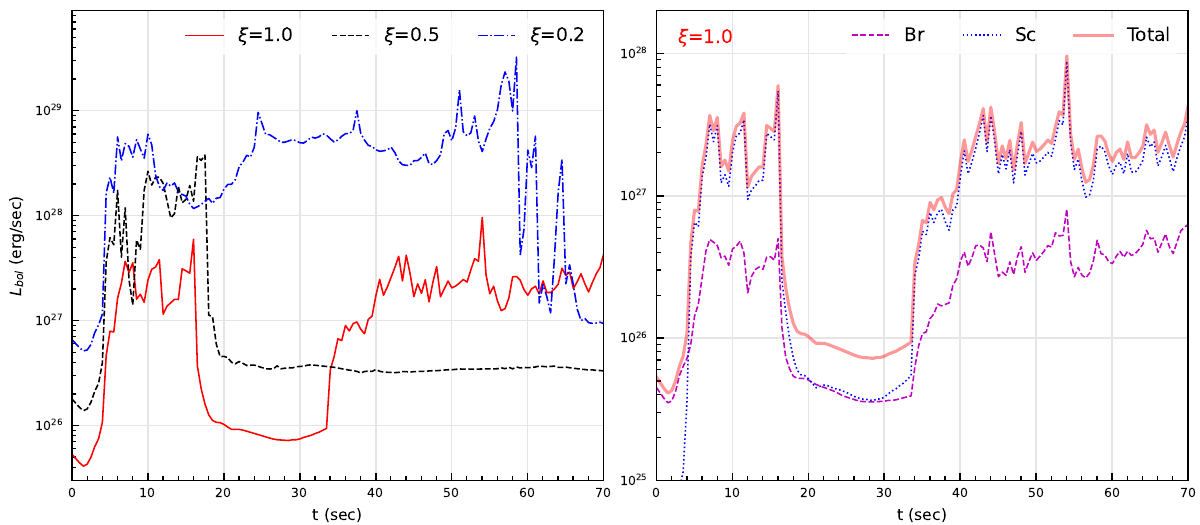}
       \caption{Variation of computed bolometric luminosity with time for models Mp5X1 ($\xi=1.0$), Mp5Xp5 ($\xi=0.5$), and Mp5Xp2 ($\xi=0.2$) are shown in the left panel, while in the right panel, the contribution of bremsstrahlung and synchrotron are shown for the case $\xi$=1.0. The value of $M_{BH}$ and $\rho_{\infty}$ are same as in figure~\ref{fig:Eqt_accretion_rates}, and plasma-$\beta$ is taken to be 100. }
       \label{fig:Eqt_lightcurve}
   \end{figure*}
 
  Figure~\ref{fig:Eqt_lightcurve} shows the temporal evolution of bolometric luminosity for three cases of composition parameters, incorporating the contributions from bremsstrahlung and synchrotron emission. As can be seen, for $\xi$=1.0, the bolometric luminosity initially increases during the disk phase, then decreases by approximately an order of magnitude when the system transitions to the shock cone phase and subsequently increases again as the disk phase resumes. While, for $\xi$=0.5, the disk structure vanishes and does not reform as indicated by the sustained lower luminosity. In contrast, for $\xi$=0.2, the disk structure persists for a significantly longer duration.

\section{\label{sec:Summary and Discussions}Summary and Discussions}
 In this paper, we have carried out numerical studies of wind accretion onto an isolated black hole using 2D numerical simulations. We considered two scenarios: (a) when the uniform parallel wind is directed toward the black hole and (b) when the uniform wind is going past the BH. We modeled the wind directed towards the BH as the classical BHL accretion and the one in which the wind streams past the BH as the lateral BHL. We employed Eulerian numerical schemes to simulate these two scenarios. The gravity is expressed through the Paczyńsky–Wiita pseudo-potential and the thermodynamics of the gas is expressed by using the variable $\Gamma$ CR EoS.
 
 Simulations of case (a), or the classical BHL, have been performed assuming axisymmetry ($r,z$ plane), which is obvious from the problem, but case (b) has been simulated on the equatorial plane ($r,\phi$ plane.
 Case (a) has been simulated where the injected wind was supersonic, while in case (b), the injected wind stream is subsonic. It may be noted that the subsonic classical BHL simulation does not produce any shock cone, while the supersonic lateral BHL forms a disk for a short time but switches back to the shock cone structure.
 
 The main goal of this paper was to find out how to form a disk-like structure if an isolated BH is bombarded by a stream of uniform wind. Since the classical BHL is an axis-symmetric system, the section of the streaming matter tugged in by the gravity of BH is symmetric on either side of the BH. Therefore, these two streams cancel each other, forcing the matter to fall directly to the BH. If the wind is streaming past the BH and is not directed towards it, then the problem cannot be cast as a classical BHL model. We call it lateral BHL since the matter is streaming past the BH, and a part of it is attracted towards the BH. This matter is then flung around the BH to form a disk. After the disk is formed, {due to the non-axisymmetric nature of the flow, the matter is non-uniformly accreted, and the spiral pattern may be disrupted.}
 In such cases, the disk breaks up, and a shock cone develops. The shock cone undergoes a flip-flop instability; however, if the thermal energy is not too high (i.e., $\Gamma \sim 5/3$), in such cases, the thermal gradient force is not too high, and the disk is restored back as we see in case of $\xi=1.0$. But for $\xi=0.5$ (less protons), although the temperature is low, the thermal energy is comparable to the rest energy of the flow $\Gamma < 5/3$, and the disk is not restored. For $\xi=0.2$, its thermal energy is even lower; therefore, the disk persists for a significantly long time, and for no protons ($\xi=0.0$), the disk is not disrupted. There are many spiral shocks in the disk. During the disk phase, the accretion rate into the BH decreases since the matter has strong rotation, and consequently, the luminosity increases by two orders of magnitude. 
 
 Based on our study, one can conclude that an isolated black hole would accrete radially if a wind is directed directly at the black hole, while if the wind is not directed towards the BH, then a temporary accretion disk may form.

\vskip 1.0cm
The authors acknowledge the anonymous referee for fruitful suggestions which significantly imporved the quality of the paper.
\bibliography{bhl_references}{}
\bibliographystyle{aasjournal}



\end{document}